\documentclass[aps,twocolumn,showpacs,superscriptaddress]{revtex4-1} 
\usepackage[english]{babel}
\usepackage{graphicx,dcolumn,bm,amssymb,amsmath,amsfonts,amsthm,latexsym}
\usepackage{float,subfig,slashed,hyperref}
\hypersetup{
  linktocpage  = true,
  colorlinks   = true,
  urlcolor     = red,
  linkcolor    = black,
  citecolor    = blue
}

\newcommand{\be}{\begin{equation}}
\newcommand{\ee}{\end{equation}}
\newcommand{\bea}{\begin{eqnarray}}
\newcommand{\eea}{\end{eqnarray}}
\newcommand{\ba}{\begin{array}{ccc}}
\newcommand{\ea}{\end{array}}
\newcommand{\nn}{\nonumber}

\begin{document}

\title{Gluon condensate from the Polyakov loop}

\author{J.P.~Carlomagno}
\email{carlomagno@fisica.unlp.edu.ar}
\affiliation{IFLP, CONICET $-$ Dpto.\ de F\'{\i}sica, Universidad Nacional de La Plata, C.C. 67, 1900 La Plata, Argentina}
\affiliation{CONICET, Rivadavia 1917, 1033 Buenos Aires, Argentina}
\affiliation{ICTP South American Institute for Fundamental Research, IFT-UNESP, Sao Paulo, SP Brazil  01440-070}
\author{J.C. Rojas}
\email{jurojas@ucn.cl}
\affiliation{Departamento de F\'{\i}sica, Universidad Cat\'olica del Norte, Angamos 0610, Antofagasta, Chile}    


\begin{abstract}
We estimate the temperature dependence of the gluon condensate from the Polyakov loop effective potential. It is presented how this analytic approach provides a simple picture for the electric gluon condensate around the deconfinement temperature, showing that it drops to zero in a temperature range which is in good agreement with different pure gauge lattice results.
\end{abstract}

\maketitle


\section{Introduction}
\label{intro}

Although Quantum Chromodynamics (QCD) is a first principle theory of hadron interactions, it has the drawback of being a theory where the low energy regime is not available using standard perturbative methods.
One characteristic feature of QCD is the presence of nontrivial gluon and chiral condensates in the system due to the strong interaction. 
Usually, for these situations, it is necessary to use effective models of QCD to describe the low energy physics.

Conservation of energy and momentum is a consequence of translation invariance, and holds whether or not the theory is scale invariant. 
Scale invariance implies that the trace of the energy-momentum tensor is zero. 
This trace condition is generally broken by quantum corrections on account of scale anomalies~\cite{Callan:1970ze}, and relates the trace of the energy-momentum tensor to the expectation value of the squared gluon field strength, the gluon condensate, through the renormalization group beta function.

The extension of the relationship of the gluon condensate and the trace anomaly was studied by Leutwyler~\citep{Leutwyler:1993}.
As it is well known, the energy momentum tensor at finite temperature can be separated into the zero temperature part and the finite temperature contribution. 
The zero temperature part contains all the ultraviolet infinities which determine the anomaly.
On the other side, the finite temperature part is connected to the thermodynamic contribution to the energy density.

In this letter we estimate the temperature dependence of the gluon condensate through the effective energy density of a pure gauge theory, i.e. disregarding the quark contribution. 
We work under the hypothesis that the effective potential of self interacting gluons is given by the Polyakov loop potential.

This work is organized as follows. In Sec.~\ref{PL} and Sec.~\ref{ggcond} we introduce the general formalism, including analytical results for the temperature dependence of the QCD gluon condensate. In Sec.~\ref{results} we present our analysis and results, and in Sec.~\ref{finale} we summarize our results and conclusions.


\section{Polyakov Loop}
\label{PL}

In the pure Yang-Mills theory the center symmetry plays a crucial part in the description of the confinement-deconfinement phase transition. 
The order parameter for the latter is the trace of the Polyakov line~\cite{'tHooft:1977hy,Polyakov:1978vu}
\bea
\Phi(x)=\frac{1}{N_c}{\rm Tr}\ {\cal P}\,\exp\left(i\int_0^{1/T}\!dx_4\,A_4\right).
\label{Polyakov_line}
\eea

The Polyakov line is not invariant under gauge transformations belonging to the gauge group center. 
Then, if $\Phi =0$ the $Z(N)$ symmetry is manifest, this situation describes confinement. 
If for any reason $\Phi \neq 0$, the symmetry must have been broken, this corresponds to the deconfined phase.
Therefore, the order parameter $\Phi$ is zero in the confined phase below the critical temperature, and assumes a non-zero value in the deconfined phase above this critical temperature. 

For our explicit calculations we shall use the freedom to rotate the $A_4$ field to a diagonal form and consider it to be static. 
In this gauge, $\Phi(x)$ is a diagonal matrix $\exp(iA_4/T)$, with $A_4$ a constant background field $A_4 = i A_0 = i g\,\delta_{\mu 0}\, A^\mu_a \lambda^a/2$, where $A^\mu_a$ are the SU(3) color gauge fields. 
Then the traced Polyakov loop (PL) is given by $\Phi=\frac{1}{3} {\rm Tr}\, \exp( i A_4/T)$. 
We will work in the so-called Polyakov gauge, in which the matrix $\Phi = \phi_3 \lambda_3 + \phi_8 \lambda_8$ is diagonal~\cite{Diakonov:2004kc}.
Owing to the charge conjugation properties of the QCD Lagrangian~\cite{Dumitru:2005ng}, the mean field traced Polyakov loop field is expected to be a real quantity. 
Assuming that $\phi_3$ and $\phi_8$ are real-valued, $\phi_8$ has to be zero~\cite{Roessner:2006xn}, and therefore
\bea
 \Phi = \frac{1}{3} [1+ 2 \cos(\phi_3/T)] \ .
\eea

The effective gauge field self-interactions are given by the Polyakov loop potential ${\cal U}\,[A(x)]$. 
At finite temperature $T$, it is usual to take for this potential a functional form based on properties of pure gauge QCD.
One possible Ansatz is that based on the logarithmic expression of the Haar measure associated with the SU(3) color group integration. 
The corresponding potential is given by~\cite{Roessner:2006xn}
\bea
\frac{{\cal{U}}_{\rm log}(\Phi ,T)}{T^4}  &&\ = 
\ -\,\frac{1}{2}\, a(T)\,\Phi^2 \;+ \nn \\
&&\;b(T)\, \log\left(1 - 6\, \Phi^2 + 8\, \Phi^3
- 3\, \Phi^4 \right) \ ,
\label{ulog}
\eea
where
\bea
&& a(T) = a_0 +a_1 \left(\dfrac{T_0}{T}\right) + a_2\left(\dfrac{T_0}{T}\right)^2 \ , \nn \\
&& b(T) = b_3\left(\dfrac{T_0}{T}\right)^3 \ . \nn
\label{log}
\eea
The parameters can be fitted to pure gauge lattice QCD data so as to properly reproduce the corresponding equation of state and PL behavior. 
This leads to~\cite{Roessner:2006xn}
\be
a_0 = 3.51\ ,\quad a_1 = -2.47\ ,\quad a_2 = 15.2\ ,\quad b_3 = -1.75\ . \nn
\ee
The values of $a_i$ and $b_i$ are constrained by the condition of reaching the Stefan-Boltzmann limit at $T \rightarrow \infty$ and by imposing the presence of a first order phase transition at $T_0$.
In absence of dynamical quarks, from lattice calculations one expects a deconfinement temperature $T_0 = 270$~MeV.
However, in the presence of light dynamical quarks this temperature scale should be adequately reduced to about 200~MeV, with an uncertainty of about 30 MeV~\cite{Schaefer:2007pw}. 

Another widely used potential is that given by a polynomial function based on a Ginzburg-Landau Ansatz~\cite{Scavenius:2002ru,Ratti:2005jh}:
\bea
\frac{{\cal{U}}_{\rm poly}(\Phi ,T)}{T ^4} \ = \ -\,\frac{b_2(T)}{2}\, \Phi^2
-\,\frac{b_3}{3}\, \Phi^3 +\,\frac{b_4}{4}\, \Phi^4 \ ,
\label{upoly}
\eea
where
\bea
b_2(T) = a_0 +a_1 \left(\dfrac{T_0}{T}\right) + a_2\left(\dfrac{T_0}{T}\right)^2
+ a_3\left(\dfrac{T_0}{T}\right)^3\ , \nn
\label{pol}
\eea
with
\bea
&a_0 = 6.75\ ,\quad a_1 = -1.95\ ,\quad a_2 = 2.625\ ,\quad a_3 = -7.44\ , \nn \\
&b_3 = 0.75\ ,\quad b_4 = 7.5\ . \nn
\eea

Here the reference temperature $T_0$ plays the same role as in the logarithmic potential of Eq.~(\ref{ulog}). 
Once again, the parameters can be fitted to pure gauge lattice QCD results so as to reproduce the corresponding equation of state and Polyakov loop behavior. 

In addition, other considered form is the PL potential proposed by Fukushima~\cite{Fukushima:2003fw,Fukushima:2008wg}, which includes both a logarithmic piece and a quadratic term with a coefficient that falls exponentially with the temperature:
\bea
{\cal{U}}_{\rm Fuku}(\Phi ,T) &&\ = \ -\,b\,T \left[\, 54\, \exp (-a/T)\, \Phi^2 \, + \right. \nn \\
&&\left. \log \left(1 - 6\, \Phi^2 + 8\, \Phi^3 - 3\, \Phi^4 \right)\, \right] \ .
\label{ufuku}
\eea
Values of dimensionful parameters $a$ and $b$ are given in Ref.~\cite{Fukushima:2008wg}. 


\section{Gluon Condensate}
\label{ggcond}

We start by considering the pure gauge QCD Lagrangian
\be
{\cal{L}}= - \frac{1}{4 g_s^2} G^a_{\mu \nu} G^{a\;\mu\nu} \ ,
\ee
where 
\be
G^{a}_{\mu\nu}=\partial_\mu A^a_\nu - \partial_\nu A^a_\mu - f^a_{bc} A^b_\mu A^c_\nu
\ee
is the gauge invariant gluon field strength tensor, where $A^a_\mu$ are the color gauge fields and $f_{abc}$ the totally antisymmetric structure constants of $SU(3)$. The strong coupling constant $g_s$ was absorbed in the gluon field, and it can be restored replacing $A_\mu$ by $g_s A_\mu$, with $g_s^2=4\pi\ \alpha_s$.

Following a decomposition sugested in Ref.~\citep{Agasian:2016hcc}, we have
\bea
\mathcal{Z} &=& \int \left[d\tilde{A} \right] \exp\left\{\ {\int \!\!\!\!\!\!\!\!\!\sum \ } {\cal L}_{eff}\right\} \nonumber \\
&\approx & \int \left[d\tilde{A} \right] \exp\left\{ 
\frac{-1}{16 \pi \alpha_s(T)} \quad 
{\int \!\!\!\!\!\!\!\!\!\sum \ } \ { G^a_{\mu \nu} G^{a\;\mu\nu}}\right\} \ .
\eea

The thermal fluctuation of the gluon condensate can be formally obtained by varying the grand canonical potential $\cal{U}$, defined as
\bea
\mathcal{U} = -T \ln \mathcal{Z} \ ,
\eea
with respect to the inverse of the coupling constant $\alpha_s$, this is
\bea
\langle G^a_{\mu \nu} G^{a\;\mu\nu} \rangle_{T}  =
16 \pi \frac{\partial\ \mathcal{U}}{\partial \alpha_s^{-1}} \ .
\eea

It has been argued that the gluon condensate can be separated in an electric and magnetic part. The later, as function of the temperature remains almost constant at its zero temperature value. While the electric contribution rapidly drops to zero near above the deconfinement temperature.
In consequence, the Polyakov loop effective potential, as an estimate for the gluons self interaction, should be related to the electric contribution of the gluon condensate. 
Accordingly, we can estimate the temperature dependence of the QCD gluon condensate $\langle G^2 \rangle$ through the thermodynamical potential ${\cal{U}}(\Phi,T)$, assuming that all the extra energy comes from the Polyakov loop effective potential. 

Restoring the coupling constant, we have
\bea
\langle \frac{\alpha_s}{\pi} G^a_{\mu \nu} G^{a\;\mu\nu} \rangle \equiv 
\langle G^2 \rangle = \langle G^2 \rangle_0 + \langle G^2 \rangle_{T,E} \ ,
\label{gluonc1}
\eea
where the value of the gluon condensate at zero temperature $\langle G^2 \rangle_0=0.037$~GeV$^4$ was taken from Ref.~\cite{Dominguez:2014pga}, and the thermal electric fluctuations are given by
\bea
\langle G^2 \rangle_{T,E} =
- \frac{4}{\pi}\alpha_{s}(T)^2 \left(\frac{\partial \alpha_s(T)}{\partial T}\right)^{-1} \frac{\partial\ \mathcal{U}(\Phi,T)}{\partial T}\ .
\label{gluonc2}
\eea

For the behavior of the strong coupling constant we choose two different parametrizations $\alpha_s^{(1,2)}(T)$ from Ref.~\cite{Deur:2016tte}. 
Here, we had assume that the temperature is the energy scale of the system, a rather valid assumption in a temperature range around $T_0$.
These coupling constants obtained through the ghost-gluon vertex, as a truncation prescription fitted in a pure gauge theory, are
\bea
\alpha_s^{(1)}(T)\  && = \frac{1}{1.16+T^4/\Lambda_1^4} 
\left[3.49\left(1.16-0.07\left(\frac{T^2}{\Lambda_1^2}\right)^{\frac{2}{3}} \right) + \right. \nn \\
&&\left. \left(\frac{T^2}{\Lambda_1^2} + 2 \right)\frac{T^4}{\Lambda_1^4} \alpha_s^{\beta_1} \right] \ ,
\label{alpha1}
\eea
where $\Lambda_1 = 0.856$~GeV, with 
\bea
\alpha_s^{\beta_1} = \frac{4 \pi}{11 \ln(T^2/\Lambda_1^2)} 
\left[1- \frac{102}{11^2} \frac{\ln(T^2/\Lambda_1^2)}{T^2/\Lambda_1^2} \right] \nn	\ .
\eea
and
\bea
\alpha_s^{(2)}(T)\ && = \frac{1}{15+T^2/\Lambda_2^2} \left[ 15\times 2.6\ + \right. \nn \\
&& \left. \frac{4\pi}{11}\left( \frac{1}{\ln(T^2/\Lambda_2^2)}-\frac{1}{T^2/\Lambda_2^2 -1} \right)\frac{T^2}{\Lambda_2^2} \right] \ ,
\label{alpha2}
\eea
with $\Lambda_2 = 0.33$~GeV.


\section{Results}
\label{results}

We present here our analytic results for the thermal fluctuations of the QCD gluon condensate defined in the previous section, considering different Polyakov loop effective potentials and parametrizations for the temperature dependence of the strong coupling constant.

\begin{figure}[h]
\includegraphics[width=0.47\textwidth]{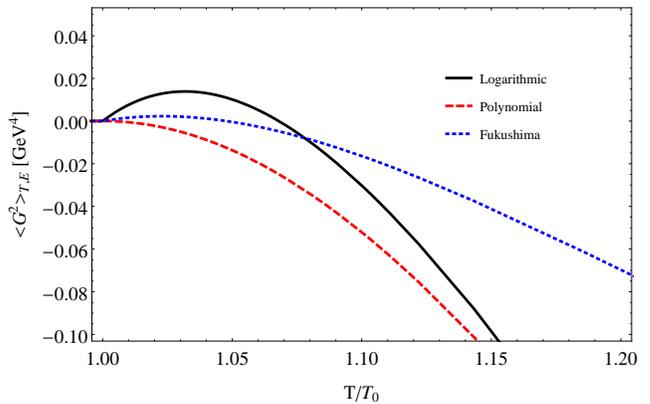}
\caption{\label{fig:1} Electric part of the gluon condensate for the PL potentials of Eqs.~(\ref{ulog}),~(\ref{upoly}) and~(\ref{ufuku}) in solid black, dashed red and dotted blue line, respectively, as functions of the reduced temperature $T/T_0$ for $\alpha_s^{(1)}(T)$.}
\end{figure}

In Fig.~\ref{fig:1} we plot for the logarithmic, polynomial and Fukushima PL effective potentials, defined in Eqs.~(\ref{ulog}),~(\ref{upoly}) and (\ref{ufuku}) in solid, dashed and dotted line, respectively, as functions of the reduced temperature $T/T_0$ the finite temperature part of the electric gluon condensate $\langle G^2 \rangle_{T,E}$, Eq.~(\ref{gluonc2}).
We set for the QCD strong coupling constant the temperature dependence of Eq.~(\ref{alpha1}).
It is worth mentioning that an equivalent behavior was found for the parametrization of Eq.~(\ref{alpha2}). 

As we are neglecting the quark contribution, we adopt $T_0 = 270$~MeV. 
Since, this value corresponds to the deconfinement transition temperature of pure gauge QCD obtained from lattice calculations~\cite{Schaefer:2007pw}.

The mean field traced Polyakov loop $\Phi$ is zero for lower temperatures than the deconfinement temperature transition $T_0$, this implies that the effective potential vanishes for those temperatures, therefore the gluon condensate remains constant at its zero temperature value $\langle G^2 \rangle_0$ up to $T/T_0 = 1$.
For higher temperatures, the gluon condensate should decreases monotonically since it is related to the free energy density. 
Hence, from Fig.~\ref{fig:1}, we see that the polynomial effective potential has the  best thermal behavior.
Consequently, we will use this potential for the estimations of the QCD gluon condensate.

\begin{figure}[h]
\includegraphics[width=0.47\textwidth]{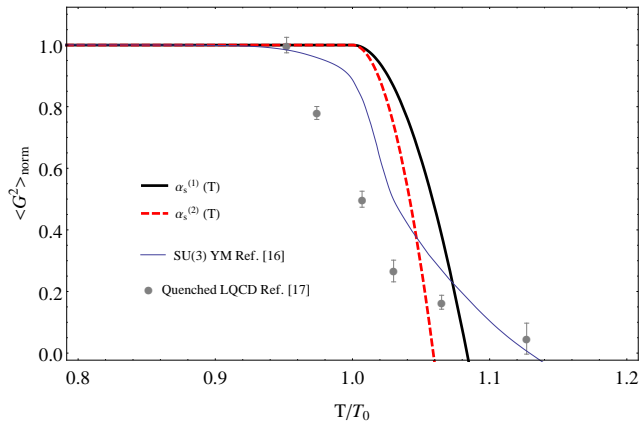}
\caption{\label{fig:2} Normalized electric gluon condensate as a function of the reduced temperature for the polynomial potential and the coupling constant $\alpha_s^{(1)}(T)$ ($\alpha_s^{(2)}(T)$) in solid black (dashed red) line. Grey circles and blue line correspond to lattice results from Ref.~\cite{DElia:2002hkf} and \cite{Boyd:1996ex}, respectively. }
\end{figure}

To compare our results, in Fig.~\ref{fig:2}, we plot Eq.~(\ref{gluonc1}) with $T_0=270$~MeV for the polynomial effective PL potential against lattice calculations for a pure gauge theory.
The grey circles were taken from Ref.~\cite{DElia:2002hkf} and correspond to a quenched lattice gauge theory. 
There, the authors show that only the electric part of the gluon condensate has a temperature dependence. 
While the blue line was obtained from Ref.~\cite{Boyd:1996ex}, here the analysis was done for a $SU(3)$ lattice gauge theory using data from Monte Carlo simulations of the interaction measure. 
The qualitative structure of their results is the same as ours, in both cases the gluon condensate drops rapidly around $T_0$ and vanishes at higher temperatures. 

Through finite energy sum rules~\cite{Ayala:2016vnt}, with inputs from lattice QCD~\cite{Dominguez:2012bs} or Nambu$-$Jona-Lasinio models~\cite{Carlomagno:2016bpu} it is possible to obtain from the dimension $4$ term in the operator product expansion, the temperature dependence of the gluon condensate via the thermal behavior of the continuum threshold.
These estimations also shown that the QCD gluon condensate remains constant up to near the transition temperature and then decreases monotonically. 
Nevertheless, this formalism provides predictions less accurate than our simple analytic approach.

Finally, as proof of consistency, we can check the parameterizations for the strong coupling constant, Eqs.~(\ref{alpha1}) and~(\ref{alpha2}). 

Previous integration of Eq.~(\ref{gluonc1}) and using lattice results from Ref.~\cite{DElia:2002hkf} for the electric gluon condensate, it is possible to obtain an approximate temperature dependence for $\alpha_s$ as a function of the PL effective potential,
\bea
\alpha_s (T) = \frac{\langle G^2 \rangle - \langle G^2 \rangle_0}{\frac{4}{\pi}\mathcal{U}(\Phi,T)+C} \ ,
\label{alphaDelia}
\eea
where $C$ is the integration constant, that could be fixed by normalization.

\begin{figure}[h]
\includegraphics[width=0.47\textwidth]{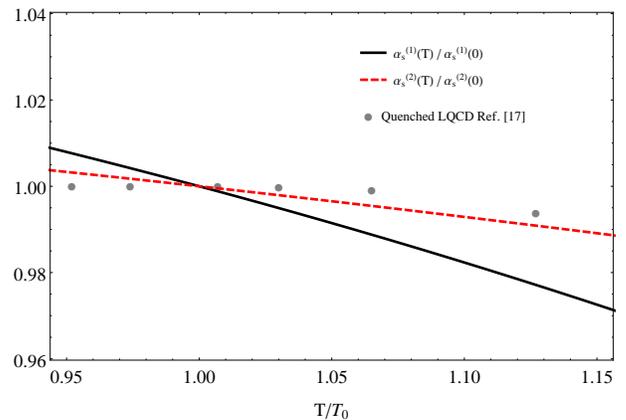}
\caption{\label{fig:3} Strong coupling constant $\alpha_s^{(1)}$ ($\alpha_s^{(2)}$) in solid black (dashed red) line as functions of $T/T_0$ and estimations from Eq.~(\ref{alphaDelia}) in grey circles obtained from Ref.~\cite{DElia:2002hkf}.}
\end{figure}

In Fig.~\ref{fig:3} we plot the two parameterizations $\alpha_s^{(1)}$ and $\alpha_s^{(2)}$ in solid black and dashed red line, respectively, against the approximate strong coupling constant of Eq.~(\ref{alphaDelia}) for the polynomial potential (grey circles), normalized by its zero temperature value as functions of the reduced temperature $T/T_0$.
We can see that booth parametrizations are in good agreement with the results from Eq.~(\ref{alphaDelia}) in the temperature range of interest.

\section{Summary and conclusions}
\label{finale}

In this work we have presented an analytic formalism without free parameters for the thermal fluctuations of the QCD gluon condensate.
It was seen that the Polyakov loop effective potential it is related to the electric part of the gluon condensate. 

We have analyzed the temperature dependence of the gluon condensate for the three most used PL potentials in the literature, and proposed two parametrizations for the temperature dependence of the strong coupling constant.

We showed that the polynomial effective potential has the best thermal behavior, while the qualitative dependence for both of the strong coupling parametrizations is equivalent.

Finally, we conclude saying that the predictions obtained within this approach are in better agreement with estimates from lattice QCD than other more complex effective theories, and provides an accurate simple picture for the description of the electric gluon condensate around the deconfinement critical temperature.


\section*{Acknowledgements}

This work has been partially funded by CONICET under Grant No.\ PIP 449; by the National University of La Plata, Project No.\ X718; by FAPESP grant 2016/01343-7 and by FONDECYT under grants No.1150471, No 1150847 and No 1170107.

The authors would like to thank M. Loewe for useful discussions.




\begin{thebibliography}{99}
\footnotesize

\bibitem{Callan:1970ze}
  C.~G.~Callan, Jr., S.~R.~Coleman and R.~Jackiw,
  Annals Phys.\  {\bf 59} (1970) 42.
  doi:10.1016/0003-4916(70)90394-5

\bibitem{Leutwyler:1993} 
H. Leutwyler, ``Deconfinement and Chiral Symmetry in QCD 20 Years Later'', Vol. 2, P. M. Zerwas and H. A. Kastrup (Eds.), World Scientific, Singapore, 1993, pp. 693-716.

\bibitem{'tHooft:1977hy} 
  G.~'t Hooft,
  Nucl.\ Phys.\ B {\bf 138}, 1 (1978).
  doi:10.1016/0550-3213(78)90153-0
  
\bibitem{Polyakov:1978vu} 
  A.~M.~Polyakov,
  Phys.\ Lett.\  {\bf 72B}, 477 (1978).
  doi:10.1016/0370-2693(78)90737-2
  
\bibitem{Diakonov:2004kc} 
  D.~Diakonov and M.~Oswald,
  Phys.\ Rev.\ D {\bf 70}, 105016 (2004)
  doi:10.1103/PhysRevD.70.105016
  [hep-ph/0403108].


\bibitem{Dumitru:2005ng} 
  A.~Dumitru, R.~D.~Pisarski and D.~Zschiesche,
  Phys.\ Rev.\ D {\bf 72}, 065008 (2005)
  doi:10.1103/PhysRevD.72.065008
  [hep-ph/0505256].
  
\bibitem{Roessner:2006xn} 
  S.~Roessner, C.~Ratti and W.~Weise,
  Phys.\ Rev.\ D {\bf 75}, 034007 (2007)
  doi:10.1103/PhysRevD.75.034007
  [hep-ph/0609281].

\bibitem{Schaefer:2007pw} 
  B.~J.~Schaefer, J.~M.~Pawlowski and J.~Wambach,
  Phys.\ Rev.\ D {\bf 76}, 074023 (2007)
  doi:10.1103/PhysRevD.76.074023
  [arXiv:0704.3234 [hep-ph]].

\bibitem{Scavenius:2002ru} 
  O.~Scavenius, A.~Dumitru and J.~T.~Lenaghan,
  Phys.\ Rev.\ C {\bf 66}, 034903 (2002)
  doi:10.1103/PhysRevC.66.034903
  [hep-ph/0201079].

\bibitem{Ratti:2005jh} 
  C.~Ratti, M.~A.~Thaler and W.~Weise,
  Phys.\ Rev.\ D {\bf 73}, 014019 (2006)
  doi:10.1103/PhysRevD.73.014019
  [hep-ph/0506234].

\bibitem{Fukushima:2003fw} 
  K.~Fukushima,
  Phys.\ Lett.\ B {\bf 591}, 277 (2004)
  doi:10.1016/j.physletb.2004.04.027
  [hep-ph/0310121].
 
\bibitem{Fukushima:2008wg} 
  K.~Fukushima,
  Phys.\ Rev.\ D {\bf 77}, 114028 (2008)
  Erratum: [Phys.\ Rev.\ D {\bf 78}, 039902 (2008)]
  doi:10.1103/PhysRevD.77.114028, 10.1103/PhysRevD.78.039902
  [arXiv:0803.3318 [hep-ph]].

\bibitem{Agasian:2016hcc}
  N.~O.~Agasian,
  JETP Lett.\  {\bf 104} (2016) no.2,  71
   [Pisma Zh.\ Eksp.\ Teor.\ Fiz.\  {\bf 104} (2016) no.2,  71]
  doi:10.1134/S002136401614006X
  [arXiv:1608.06773 [hep-ph]].

\bibitem{Dominguez:2014pga} 
  C.~A.~Dominguez, L.~A.~Hernandez and K.~Schilcher,
  JHEP {\bf 1507}, 110 (2015)
  doi:10.1007/JHEP07(2015)110
  [arXiv:1411.4500 [hep-ph]].

\bibitem{Deur:2016tte} 
  A.~Deur, S.~J.~Brodsky and G.~F.~de Teramond,
  Prog.\ Part.\ Nucl.\ Phys.\  {\bf 90}, 1 (2016)
  doi:10.1016/j.ppnp.2016.04.003
  [arXiv:1604.08082 [hep-ph]].
  
\bibitem{Levai:1997yx} 
  P.~Levai and U.~W.~Heinz,
  Phys.\ Rev.\ C {\bf 57}, 1879 (1998)
  doi:10.1103/PhysRevC.57.1879
  [hep-ph/9710463].
  
\bibitem{Boyd:1996ex} 
  G.~Boyd and D.~E.~Miller,
  hep-ph/9608482.
  
\bibitem{DElia:2002hkf} 
  M.~D'Elia, A.~Di Giacomo and E.~Meggiolaro,
  Phys.\ Rev.\ D {\bf 67}, 114504 (2003)
  doi:10.1103/PhysRevD.67.114504
  [hep-lat/0205018].
  
\bibitem{Ayala:2016vnt} 
  A.~Ayala, C.~A.~Dominguez and M.~Loewe,
  Adv.\ High Energy Phys.\  {\bf 2017}, 9291623 (2017)
  doi:10.1155/2017/9291623
  [arXiv:1608.04284 [hep-ph]].

\bibitem{Dominguez:2012bs}
  C.~A.~Dominguez, M.~Loewe and Y.~Zhang,
  Phys.\ Rev.\ D {\bf 86} (2012) no.3,  034030
   Erratum: [Phys.\ Rev.\ D {\bf 90} (2014) no.3,  039903]
  doi:10.1103/PhysRevD.90.039903, 10.1103/PhysRevD.86.034030
  [arXiv:1205.3361 [hep-ph]].
  
\bibitem{Carlomagno:2016bpu} 
  J.~P.~Carlomagno and M.~Loewe,
  Phys.\ Rev.\ D {\bf 95}, no. 3, 036003 (2017)
  doi:10.1103/PhysRevD.95.036003
  [arXiv:1610.05429 [hep-ph]].

\end{thebibliography}
\end{document}